\documentclass[floatfix,twocolumn,10pt,pra,aps]{revtex4-2}
\usepackage{lipsum}
\usepackage{subfigure}
\usepackage{comment}
\usepackage[dvipsnames]{xcolor}
\usepackage{float}
\usepackage{graphicx}
\usepackage{amsmath}
\usepackage{subfigure}
\usepackage[normalem]{ulem} 
\usepackage{color}
\usepackage{natbib}
\begin{document}

\title{A prediction for 25th solar cycle using visibility graph and Hathaway function}

\author{Eduardo Fl\'andez$^1$}
		\email{eduardo.flandez@ug.uchile.cl}
\affiliation{{$^1$}Departamento de F\'isica, Facultad de Ciencias, Universidad de Chile, Casilla 653, Santiago, Chile}

\author{V\'{\i}ctor Mu\~noz$^1$}
	%\email{vmunoz@macul.ciencias.uchile.cl}
\affiliation{{$^1$}Departamento de F\'isica, Facultad de Ciencias, Universidad de Chile, Casilla 653, Santiago, Chile}

\begin{abstract} 
We apply a complex network approach to analyse the time series of
five solar parameters, and propose an strategy to predict the number
of sunspots for the next solar maximum, and when will this maximum
will occur. The approach is based on the Visibility Graph (VG)
algorithm, and a slightly modified version of it, the Horizontal
Visibility Graph (HVG), which map a time series into a complex
network. 
Various network metrics exhibit either an exponential or a scale-free
behavior, and we find that the evolution of the 
characteristic decay exponents is consistent with variations of the
sunspots number 
along solar cycles. During solar minimum, the sunspots
number and the solar index time series 
have characteristic decay exponents that
correlate well with the next maximum sunspots number, suggesting
that they may be good precursors of the intensity of the next solar
maximum. Based on this observation, we find that, based on current
data, the algorithm predicts a number of 179 sunspots
for cycle 25. Combining this with the Hathaway function, adjusted to
yield such maximum sunspots number, we find that the maximum for solar
cycle 25 will occur in December 2024/January 2025. 
\end{abstract}

\maketitle

\section{Introduction}
\label{introduction}

The number of sunspots is one of the main signatures of the variation
of solar magnetic activity, having been recorded for centuries. It
shows that solar activity exhibits an 11-year cycle, where the maximum
of the cycle has the largest number of sunspots and solar
flares~\citep{schrijver2009heliophysics,sturrock2013physics,schrijver2010heliophysics},
but although the most accepted explanation to date for the origin and
dynamics of the solar magnetism is the dynamo
theory~\citep{charbonneau2014solar,
  aschwanden2006physics,kamide2007handbook,lang2008sun,schrijver2009heliophysics},
modeling of the solar cycle remains a challenge. In particular
regarding some specific features of the solar cycle: when will the
next solar maximum occur, when will it start, how long it will last,
and the maximum sunspots number, given their relevance to space weather
parameters such as solar flare frequency and intensity, solar wind
conditions, and geomagnetic activity.

Forecasting models based on concepts from nonlinear dynamics theory
applied to experimental time series, such as embedding phase space,
Lyapunov spectrum, and chaotic behavior, have been
proposed~\citep{sello2001solar}. There are also surface flux transport
(SFT) models, that describe the transport of magnetic flux across 
the solar surface, modelling it as an advective-diffusive
transport~\citep{wang2017surface, jiang2014magnetic},  and models
based on solar dynamo, based on a mean-field theory approach, where
a system of coupled partial differential equations governs the
evolution of the toroidal and poloidal components of the large-scale
magnetic field~\citep{cameron2017global, charbonneau2010dynamo}.

Also, various methods have been proposed to predict the solar
cycle (see {\it e.g.\/}~\citep{hathaway1999synthesis}), based on
statistical correlations between solar and/or geomagnetic quantities. 

The essential idea is that there exist historical correlations between
solar activity indices during cycle minima and the following
maxima.~\citep{petrovay2020solar}
If such correlations are found, then observing the sun
during a certain minimum can provide information to forecast features
of the next cycle. 
For instance, a regression technique has been proposed for predicting
solar activity levels one year ahead~\citep{mcnish1949prediction}, and
functions which fit the sunspots number curve, based on two free
parameters for each solar cycle: a starting time and an
amplitude~\citep{hathaway1994shape}. Varios quantities have been
proposed to use these correlations, in sunspot number
records~\citep{wilson1998estimate}, in the Sun's polar magnetic field,
and in geomagnetic indices~\citep{feynman1982geomagnetic}.

% It may perhaps be curious that a geomagnetic index serves as a
% precursor to solar activity. 
% The correlation emerges because the geomagnetic indices mirror the strength of the poloidal field, which becomes the prevailing component of solar magnetism during activity minimum. Additionally, the dynamo theory posits that throughout a cycle, solar magnetism alternates between poloidal and toroidal components, with a net field amplification during the transition from poloidal to toroidal, as suggested by the authors of Ref.~\citep{schatten1978using}.

Recently, the visibility graph technique has shown to be an
interesing tool for time series analysis, as it provides
a method to transform a time series into a complex network, whose
metrics can be used to explore the statistical properties of the time
series~\cite{lacasa2008time}. 
It has been used to characterize the light curves of pulsating
variable stars~\citep{munoz2021analysis}, to study variations in solar
activity along solar cycles ~\citep{zou2014long,
  zurita2023characterizing}, to analyze solar wind
velocities~\citep{suyal2014visibility}, and to study the statistical
properties of solar 
flares~\citep{yu2012multifractal, taran2022complex}. These works have
shown that the metrics of the visibility graphs can be correlated to
various stages of solar magnetic activity along the solar cycle.

Following these ideas, we propose that visibility graphs can provide a
further means to find correlations between features of the solar
minima and the next solar maxima, in the spirit of similar strategies
based on the time series
itself~\citep{hathaway1999synthesis,petrovay2020solar,hathaway2015solar}.

The paper is structured as follows. In Section \ref{data}, the datasets utilized for the analysis are outlined.
 Then, in Sec.~\ref{method} the
relevant complex network and visibility graph concepts are presented.
Results are shown in Sec.~\ref{results}, and they are discussed and
summarized in Sec.~\ref{summary}.
\section{Data}\label{data}

\begin{figure}[H]
	\centering
	\includegraphics[height =0.35\textheight]{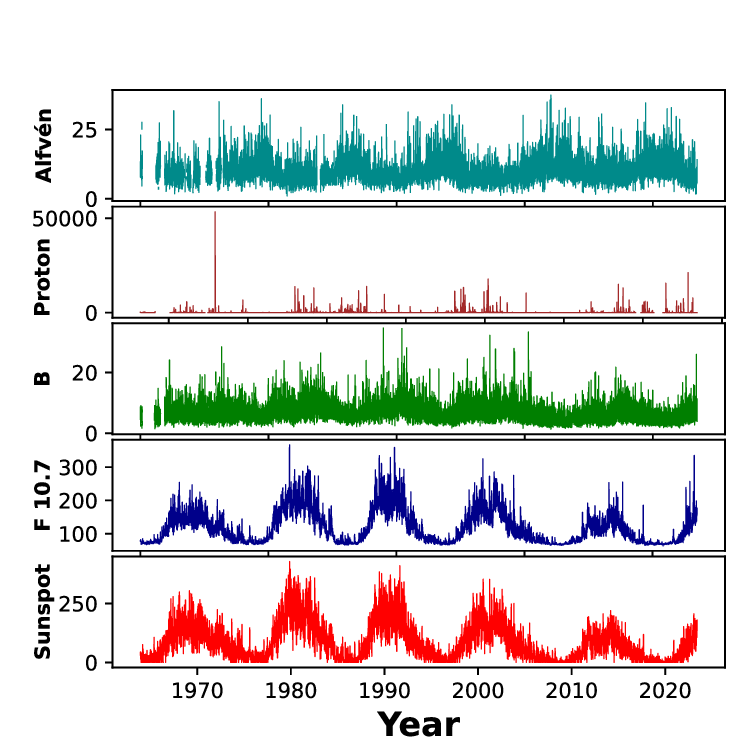}
	\caption{Daily values for Alfv\'en Mach number, proton flux,
          solar magnetic field, F10.7 index and sunspot numbers,
          corresponding to solar cycles 20--24, and the beginning of~25.}
	\label{fig:parameters}
\end{figure}

We build complex networks using the time series of daily data for five parameters
related to solar magnetic activity, from solar cycle 20 until June 11, 2023,
 corresponding to the ascending phase of SC 25. The chosen parameters are:
Alfv\'en Mach number, proton flux, magnetic field, 10.7~$\mathrm{cm}$ radio
flux and sunspots number, consistent with Ref.~\citep{singh2017early},
in order to allow to compare with previous research. 
The
respective time 
series with a cadence of one day are plotted in
Fig.~\ref{fig:parameters}.

Data have been taken from NASA 
Space Physics Data Facility (SPDF)
({\tt https://omniweb.gsfc.nasa.gov/ow.html}). 
The Low Resolution OMNI (LRO) dataset is primarily a compilation of hourly-averaged, near-Earth solar wind magnetic field and plasma parameter data from various spacecraft in geocentric or Lagrange point 1 (L1) orbits, spanning from 1963 to the present
(\url{https://omniweb.gsfc.nasa.gov/html/ow_data.html}). Additional
details about the Alv\'en Mach number, proton flux, and the F10.7
index are given below.

The Alfv\'en Mach number ($M_A$) is the ratio between plasma velocity
$V$ and Alfv\'en velocity $V_A$, and has been used to model and
characterize the space
weather~\citep{maguire2020evolution,hartigan2015new}:
\begin{equation}
 M_A = \frac{V}{V_A}	\ .
\end{equation}
%The OMNI dataset lacks the advantage of individual bow shock crossing
%fit results. This is because the bow shock's speed changes rapidly
%while spacecraft are distant from it. 
In general, the shock speed
should be included in the Mach number calculation, necessitating some
additional assumptions. Since the bow shock typically remains in front
of the Earth, assuming it is stationary is reasonable for long-term
averages. With a stationary bow shock, the Mach number simplifies to
the one described above.

Solar proton events (SPEs) are more frequent during solar maximum, typically caused by rapid coronal mass ejections. During these events, satellites face increased exposure to high-energy particles~\citep{nakano1968evidence, williams1969proton}. Proton flux data from various spacecraft, such as IMP and GOES, is available in OMNI and OMNI 2 databases, covering energies above key thresholds like 1, 10, and 60 MeV, with measurements spanning several decades from different missions (\url{https://satdat.ngdc.noaa.gov/sem/goes/data/avg}).
Detailed information about the instrument and data can be found at \url{http://sd-www.jhuapl.edu/IMP/imp_index.html} and \url{http://imp.ftecs.com}.

The solar spectral irradiance at 10.7 cm (F10.7, expressed in units of $10^{-22} \mathrm{ W {m}^{-2} {Hz}^{-1}}$) is used as an index of solar activity. Each F10.7 value represents the total emission at a wavelength of 10.7 $\mathrm{cm}$ from all sources present in the solar disk, measured over a period of one hour. Wavelengths in the 10 cm region are particularly effective for monitoring solar activity levels because emissions at these wavelengths are highly sensitive to conditions in the upper chromosphere and the base of the corona~\citep{tapping201310}.

Based on each time series in Fig.~\ref{fig:parameters}, we will study
their temporal variability along the four solar cycles
under consideration, in order to derive parameters that may provide
clues to predict when Solar Cycle 25 will occur. This will be achieved
using complex networks, as explained in Sec.~\ref{method}.

\section{Method}
\label{method}

As mentioned in Sec.~\ref{introduction}, some correlations between the
value of solar or geomagnetic variables during solar minimum, and the
number of sunspots at the following solar maximum have been found.
This is related to the idea that 
the magnetic field at the Sun's poles serves as a precursor for the
next cycle. It has been shown
that there is a linear correlation between the
difference in the magnitude of north and south magnetic fields during
solar minimum, $\Delta B_{\text{poles}} $, and the next maximum number of
sunspots, so that when the difference between the north and south
polar magnetic fields is higher, then the maximum number of sunspots is
also higher. 
The Sun's polar field difference
$\Delta B_{\text{poles}} $ has been one the most reliable
predictors of the last three cycles, since measurements have been
available~\cite{hathaway2015solar}. Based on the 24/25 cycle minimum in December 2019, the prediction indicates a maximum sunspot
number of 120 for cycle 25~\cite{hathaway2015solar}. 

Another example is geomagnetic activity near cycle
minima~\citep{ohl1966wolf}, which has the advantage of 
having more available data, namely for the last 13 sunspot cycles.
This involves the $aa$ geomagnetic index,
which measures the amplitude 
of geomagnetic disturbances in nanoteslas ($\mathrm{nT}$) from records
of two geomagnetic stations,  one in each of the Earth's
hemispheres. The higher the value of the aa index, the more intense
the geomagnetic disturbance. 
It has been observed that the minimum of geomagnetic activity, 
 as measured by the $aa$ index, is correlated to the amplitude of the
 next cycle~\citep{schatten1996solar}. Using data for the 24/25 cycle
 minimum, this approach leads to a maximum sunspot number of 132 for
 cycle 25~\cite{hathaway2015solar}. 

Given the intrinsic nonlinearity of solar and space physics phenomena,
there has been an 
increasing interest in using complex systems approaches to describe
and model them,~\citep{aschwanden201625, Dendy} and gain new insights
on issues such as
the occurrence of geomagnetic
storms~\citep{dominguez2014temporal} or the passage or magnetic
structures in the solar wind~\citep{Munoz_ao}. Complexity approaches
have also been 
applied to the problem of forecasting the solar cycle.
In Ref.~\citep{singh2017early}, an analysis of long memory processes in solar activity was made.  They considered the same solar parameters as in
Fig.~\ref{fig:parameters}, but for years 1976--2016. 
By applying a statistical test of persistence of solar activity using
the Hurst exponent $H$, they determined the maximum number of sunspots
to be 
102.8 $\pm$ 24.6. Through simplex projection analysis, they forecasted that solar cycle 25 would begin in January 2021 and last until September 2031, with its maximum occurring in June 2024.

In our work, we intend to establish to what
extent a complex network analysis can be used to predict features of
the next solar maximum, based on the same parameters as in
Ref.~\citep{singh2017early}.
Complex
networks have become a widely used 
strategy for studying complex systems, as they allow the abstraction
of systems made up of a large number of interrelated components in a
very general way. They have provided a useful approach to tackle 
various issues in space
physics~\citep{zou2014complex,lu2018complex},  earthquakes
analysis~\citep{abe2004scale, abe2006complex, 
  abe2011finite, pasten2017time, abe2011universalities,
  pasten2018time, pasten2018non}, atmospheric
flows~\citep{charakopoulos2018dynamics}, 
and numerous other
 research areas (see for example Ref.~\citep{albert2002statistical}).
Regarding solar
activity, complex networks have been used to 
study time reversibility in turbulent states in solar wind
simulations~\citep{acosta2021applying}, to assess the probability of
flares in solar active regions~\citep{daei2017complex}, to analyze
solar magnetograms~\citep{munoz2022complex}, characterize the sunspot
time series~\citep{zou2014complex}, and more recently, to explore solar
flare statistics~\citep{najafi2020solar, gheibi2017solar,
  taran2022complex, lotfi2020ultraviolet}.

A network consists of vertices (nodes) connected by edges.
 Physically, nodes can represent entities that constitute a given system (people in a society, molecules in a solid, servers on the Internet, etc.), and connections can represent links or interactions between them (friendship relationships, electrical interactions, flow of data, etc.). If the connections in a network have an assigned orientation, the network is said to be directed. If the connections have no orientation, the network is undirected~\citep{newman2018networks}.
For this work, we construct a complex network using the visibility
graph (VG) algorithm, which converts a time series into a graph. In
this graph, each node represents a data point in the series, and two
nodes are connected if both nodes can ``see'' each other (the straight
line which joins both points lies above all intermediate data
points~\citep{lacasa2008time}). 

We also consider a variation of the above called the horizontal
visibility algorithm (HVG), where visibility is given by horizontal
lines~\citep{luque2009horizontal}.
Visibility graphs have been shown to be an interesting approach to
extract statistical properties from time series, which can be in turn
related to the underlying physical processes leading to such time
series. For instance, it has been shown that the VG is able to
distinguish between time series generated by fractal versus fractional
brownian motion processes~\citep{lacasa2009visibility}, whereas the HVG can distinguish
uncorrelated, or correlated stochastic
processes~\citep{luque2009horizontal}. This has led
to various works which use VG/HVG techniques to characterize the
physics which leads to the observed time series, such as unveiling
correlations in earthquakes sequences~\citep{kundu2021extracting},
studying the degree of persistence in Brownian motion
processes~\citep{lacasa2009visibility}, screening of sleep
disorders~\citep{li2023convolutional}, quantifying irreversibility issues in
physical processes~\cite{lacasa2012time} (such as interplate vs.~intraplate seismicity~\citep{telesca2020analysis}), discriminating between
pulsation modes in variable stars~\citep{munoz2021analysis},
and following the evolution of solar magnetic
activity~\citep{zurita2023characterizing, zou2014complex}.

Based on these results, in this paper we study whether the VG approach
can provide metrics which can, in turn, be correlated to solar
magnetic activity and be used as precursors for the amplitude of the
next solar cycle.

VG and HVG are determined for each of the time series in
Sec.~\ref{data}. In order to define the
graph, a time window of data must be selected, and two choices will be
considered: the complete data set, and sliding windows. We construct
undirected networks, 
whose metrics are then analized. Details are
provided in Sec.~\ref{results}.

\section{Results}
\label{results}

One of the simplest measures for a network is the degree $k$ (number of 
connections per node). Despite its simplicity, the probability
distribution of the degree $P(k)$ can provide information
about the underlying processes, revealing whether it is a random
process or it follows some kind of preferential
attachment~\cite{albert2002statistical}, and various studies have shown
the usefulness of the characteristic decay exponents at the tail of
the distribution to characterize the topology of the complex network
and its 
evolution~\citep{barabasi1999emergence,munoz2021analysis,zurita2023characterizing,pasten2017time,munoz2022complex,Guillier_b,zou2014complex,munoz2021analysis}.

Thus, we build a complex network via
the VG algorithm, for each one of the five solar parameters in
Sec.~\ref{data}, 
and study its degree distribution $P(k)$. Results show that
distributions follow a power-law behavior, with characteristic
exponent $\gamma$, both for the complete time series and for 
sliding windows (11-years wide, shifted by one year with respect to
the previous one). This is consistent with previous works~\citep{lacasa2008time, munoz2021analysis}.

\begin{figure}[H]
	\centering
	\includegraphics[height =0.2\textheight]{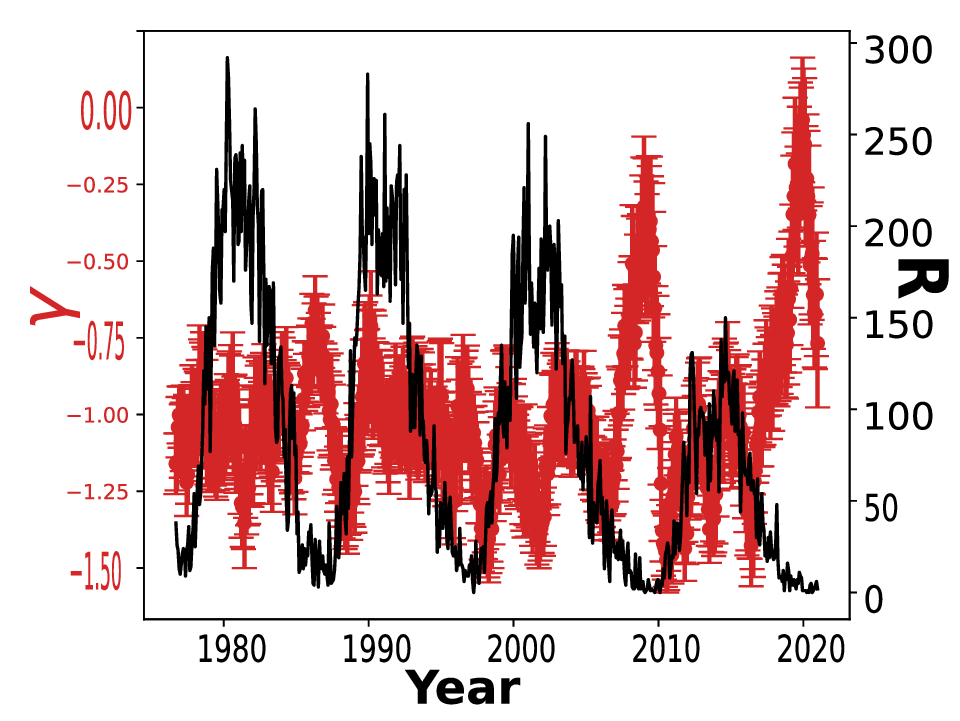}
	\caption{Decay exponent
		$\gamma$ for the degree distributions of visibility
                graphs, using
		11-years moving windows of the sunspots time series.
                Red curve: decay exponent $\gamma$; black curve:
                average sunspots number $R$ during the time window.}
	\label{fig:gamma_sunspot_sunspot}
\end{figure}

However, in the spirit of
previous works such as \cite{zou2014complex}, a plot of $\gamma$ as a
function of the mean number of 
sunspots $R$ for the corresponding time window is made, but it does not show a direct
correlation for any of the five times series, as illustrated in
Fig.~\ref{fig:gamma_sunspot_sunspot}.

\begin{figure}[H]
	\centering
	\subfigure[~Alfv\'en Mach number]{\includegraphics[width=0.48\linewidth]{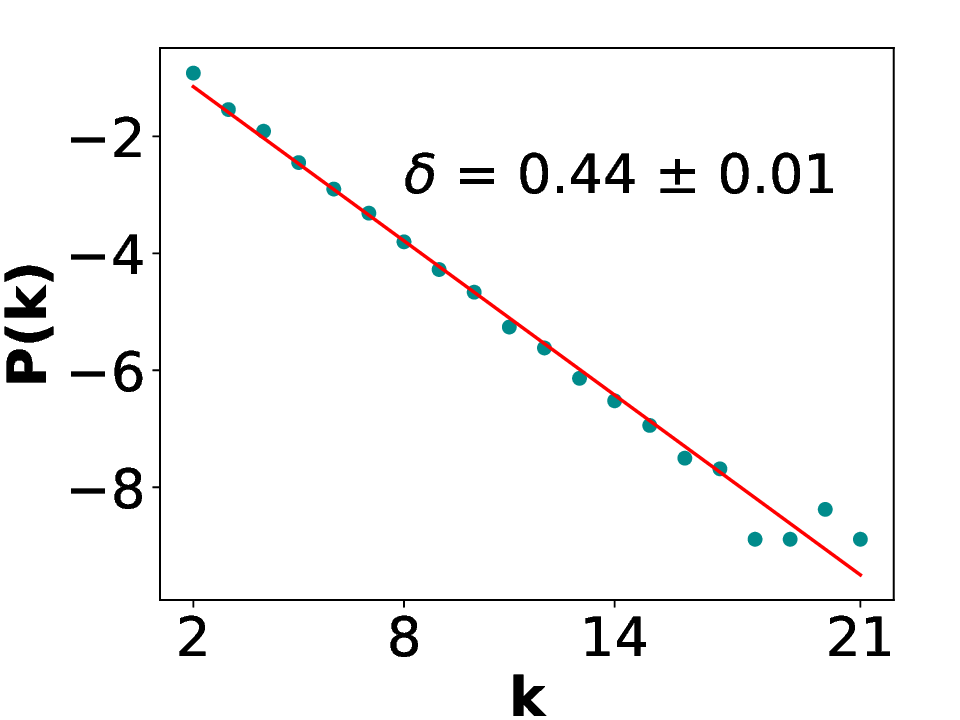}}
	\subfigure[~Proton flux.]{\includegraphics[width=0.48\linewidth]{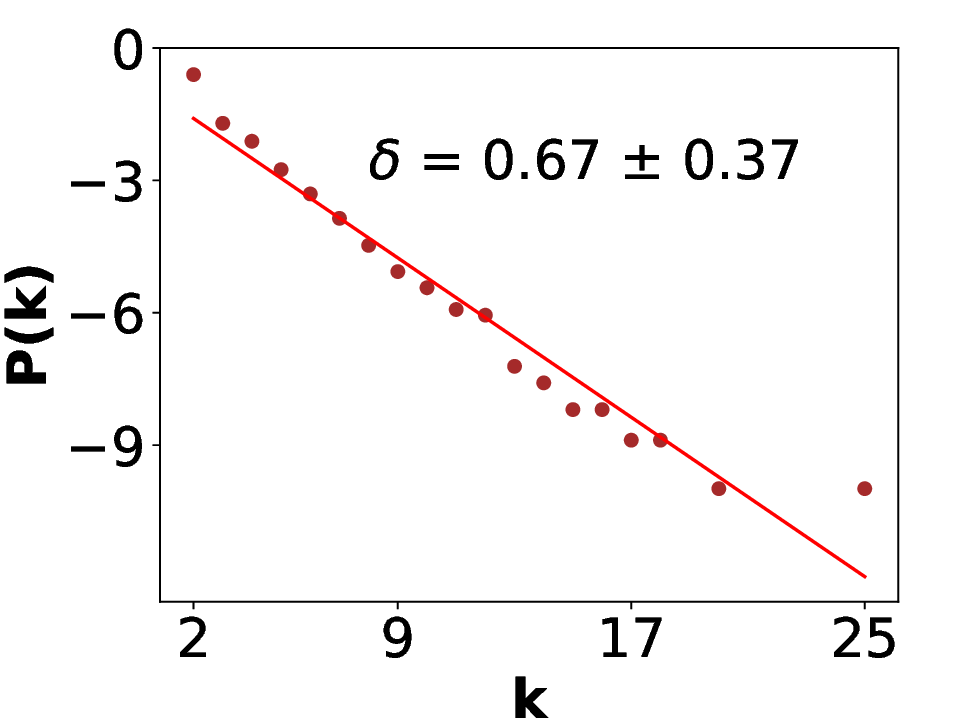}}
	\subfigure[~Solar magnetic field]{\includegraphics[width=0.48\linewidth]{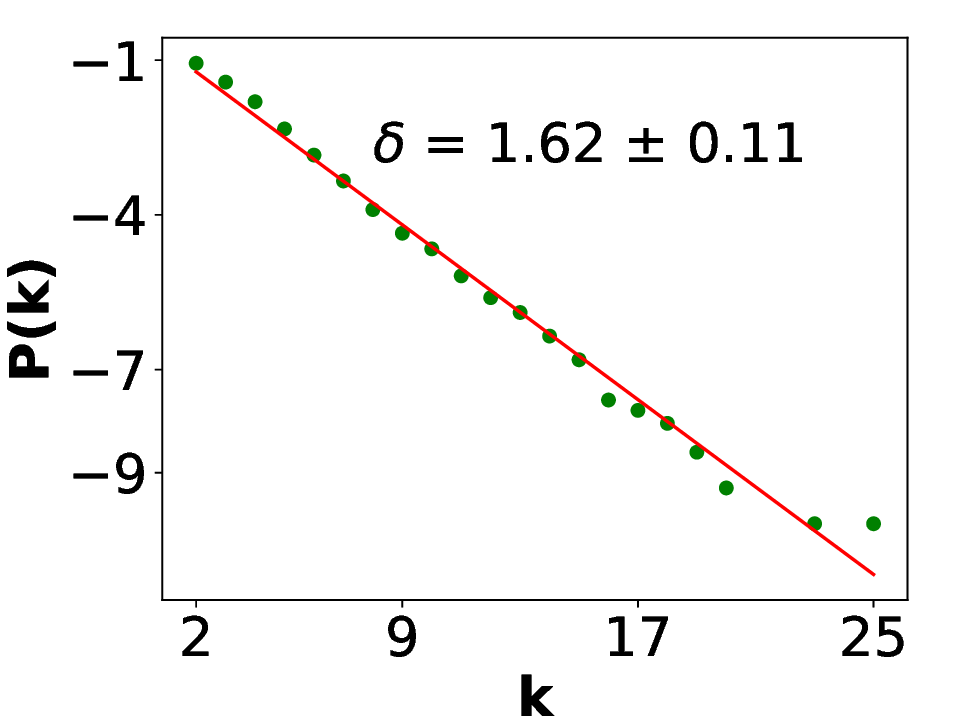}}
	\subfigure[~10.7 cm radio flux]{\includegraphics[width=0.48\linewidth]{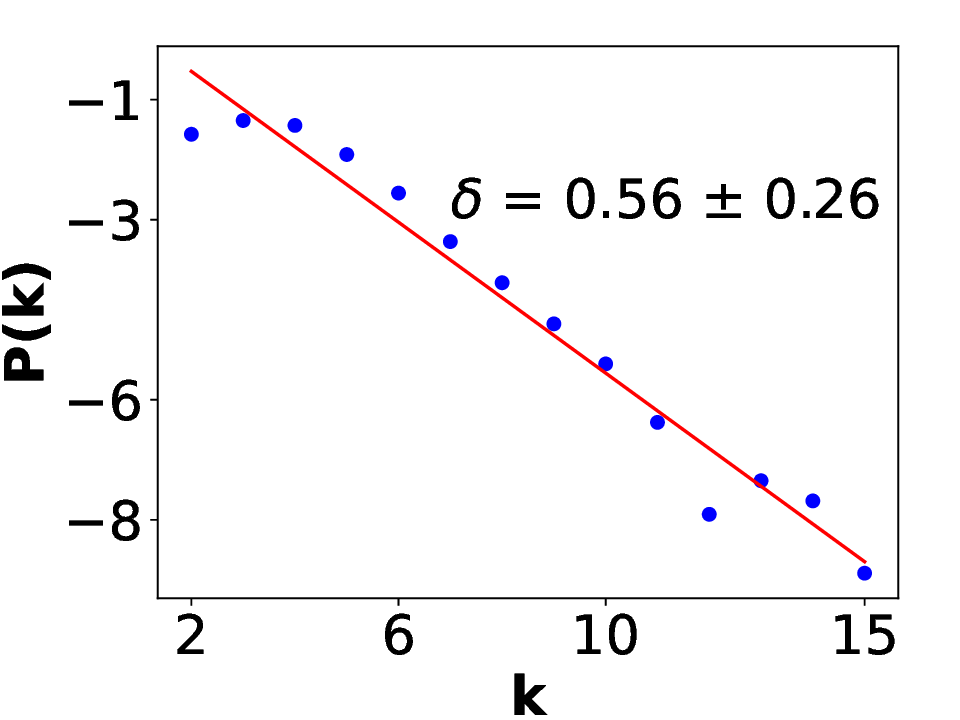}}
	\subfigure[~Sunspots]{\includegraphics[width=0.48\linewidth]{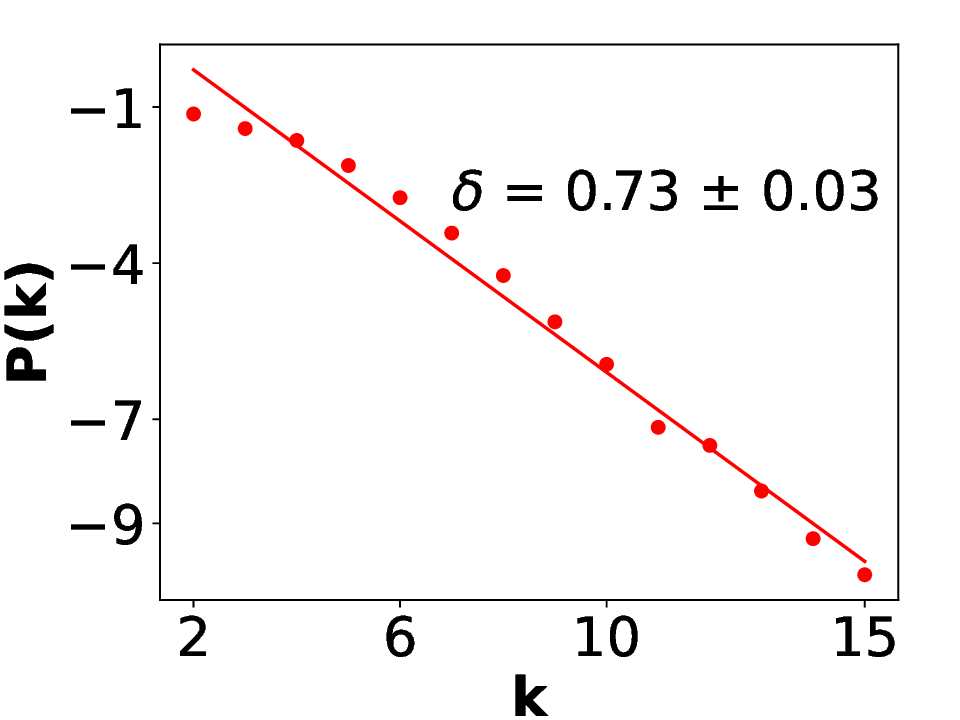}}
	\caption{Degree distributions for HVG, for each solar
          parameter in Fig.~\ref{fig:parameters}, in semilogarithmic
          scale. }  
	\label{hvg}
\end{figure}

We now repeat the previous analysis, but for the Horizontal Visibility
Graph (HVG) algorithm~\citep{luque2009horizontal}. The resulting
degree distributions for each time series are exponential, with decay
exponent $\delta$,
which is consistent with previous works~\citep{Munoz_ao,lacasa2008time,luque2009horizontal,zou2014complex,kaki2022evidence}. A semilog plot of the distributions, with their
corresponding characteristic exponents and best fits are shown in Fig.~\ref{hvg}.

The fit for the HVG
algorithm (Fig.~\ref{hvg}) turns out to be better than for the VG
algorithm, being well represented by a function 
of the form $e^{-\delta k}$. This suggests that the HVG can be more
useful than the VG in this case. The respective characteristic decay
exponents are shown in Fig.~\ref{hvg}, and in Table~\ref{table_gamma}.

\begin{table}[h]
	\centering
	\begin{tabular}{lc}
		Solar parameter & $\delta$ \\ \hline
		Alfv\'en Mach number & $ 0.44 \pm 0.01$ \\
		Proton flux & $0.67 \pm 0.37$\\
		Magnetic field & $ 1.62 \pm 0.11 $\\
		10.7~cm radio flux &  $ 0.56 \pm 0.26$\\
		Sunspots & $0.73 \pm 0.03$
	\end{tabular}
	\caption{Characteristic exponent for each time series,
          obtained from linear fit for Fig.~\ref{hvg}.}
	\label{table_gamma}
\end{table}

\begin{figure}[H]
	\centering
	\subfigure[Alfv\'en Mach number]{\includegraphics[width=0.47\linewidth]{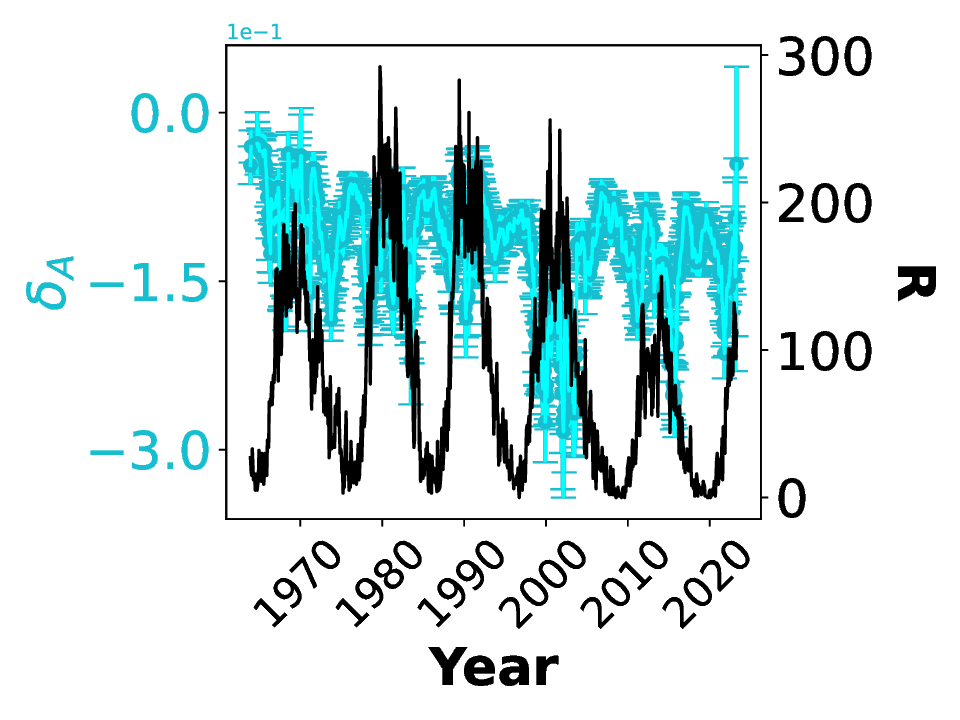}}
	\subfigure[Proton flux]{\includegraphics[width=0.47\linewidth]{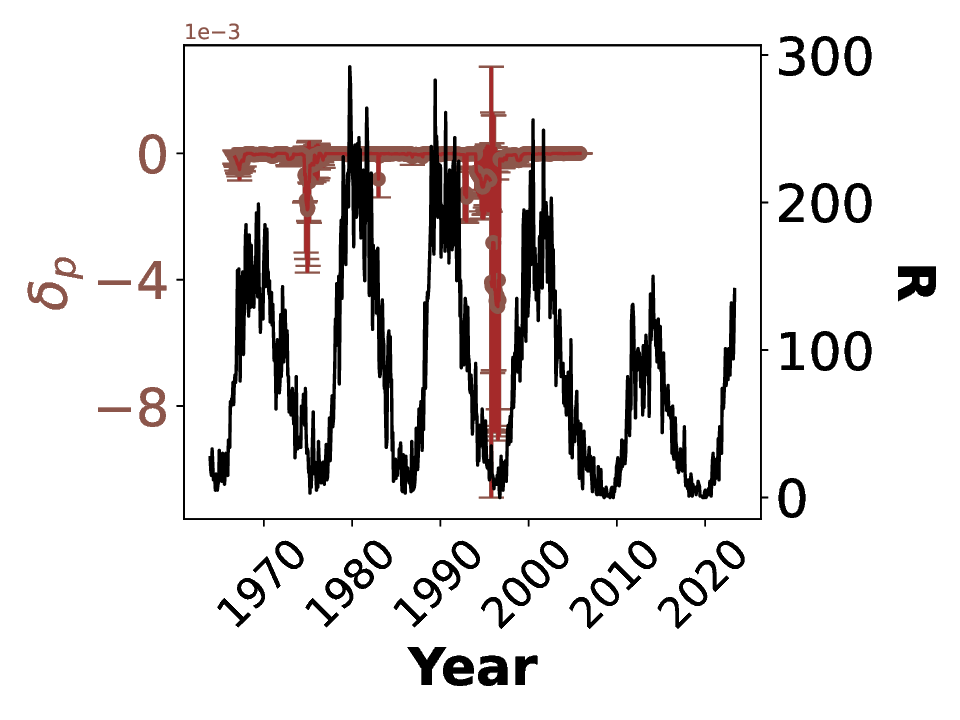}}
	\subfigure[Magnetic field]{\includegraphics[width=0.47\linewidth]{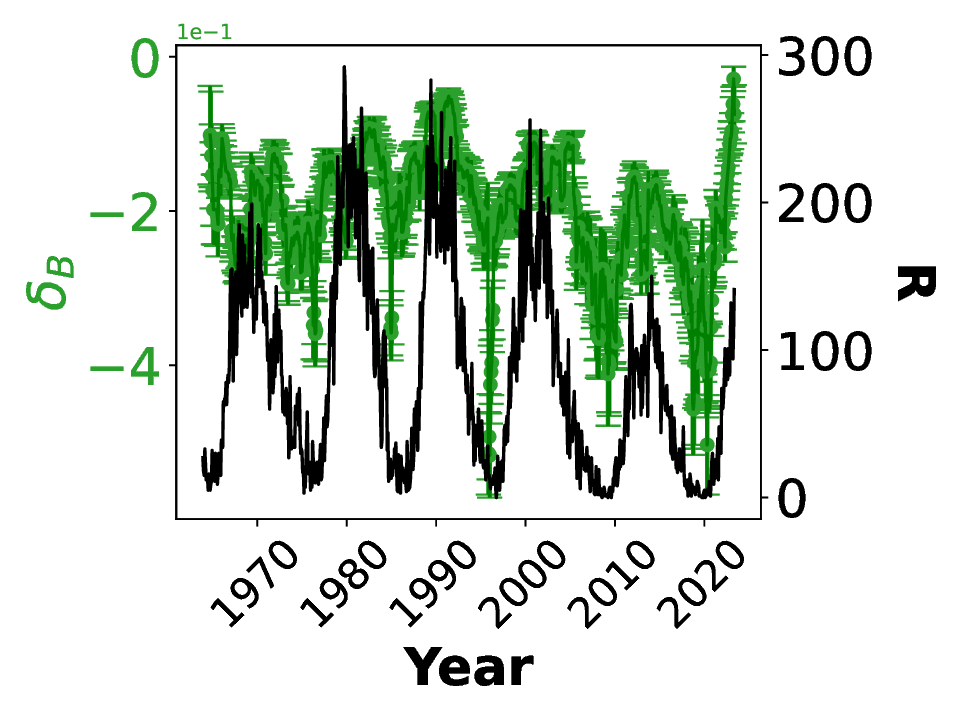}}
	\subfigure[F 10.7 index]{\includegraphics[width=0.47\linewidth]{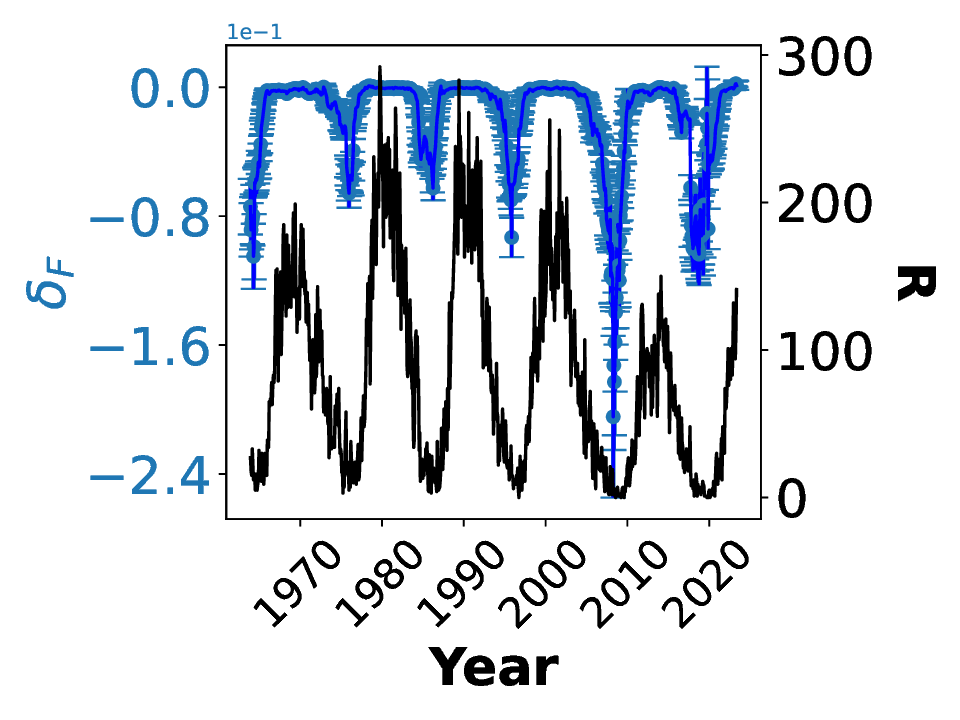}}
	\subfigure[Sunspot number]{\includegraphics[width=0.47\linewidth]{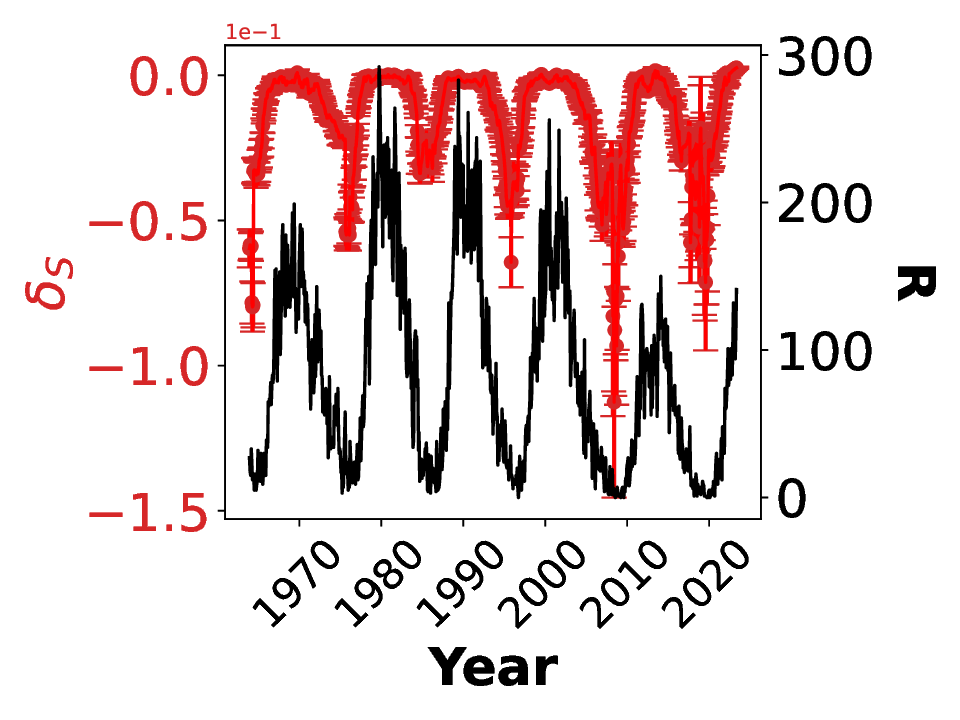}}
	\caption{Decay exponent
          $\delta$ obtained from horizontal visibility graphs for
          11-years wide time moving windows. Black curves: average sunspots
          number $R$ during the time window.}
	\label{fig:HVG}
\end{figure}

For the moving windows analysis, the plot of the decay exponent is
shown in Fig.~\ref{fig:HVG}.
In Fig.~\ref{fig:HVG}, we note that there is
a periodic variation in the exponent $\delta$ for the solar parameters
that is consistent with the variation of the solar cycles, except for
the proton flux.

The most interesting curves are found for the 10.7 cm radio
flux and the sunspots 
curves, Figs.~\ref{fig:HVG}(d) and~(e), which show a clear anticorrelation
with the sunspots number, where deeper minima in $\delta$ occur before
lower maxima of $R$. This is interesting, as it reminds similar
correlations found in previous works
on precursors based on completely different
quantities,~\citep{hathaway2015solar} 
as discussed
in Sec.~\ref{introduction}. 
Thus, we will focus on those two solar
parameters.

\begin{figure}[H]
	\centering  
	\subfigure[F 10.7 index ]{\includegraphics[width=0.45\linewidth]{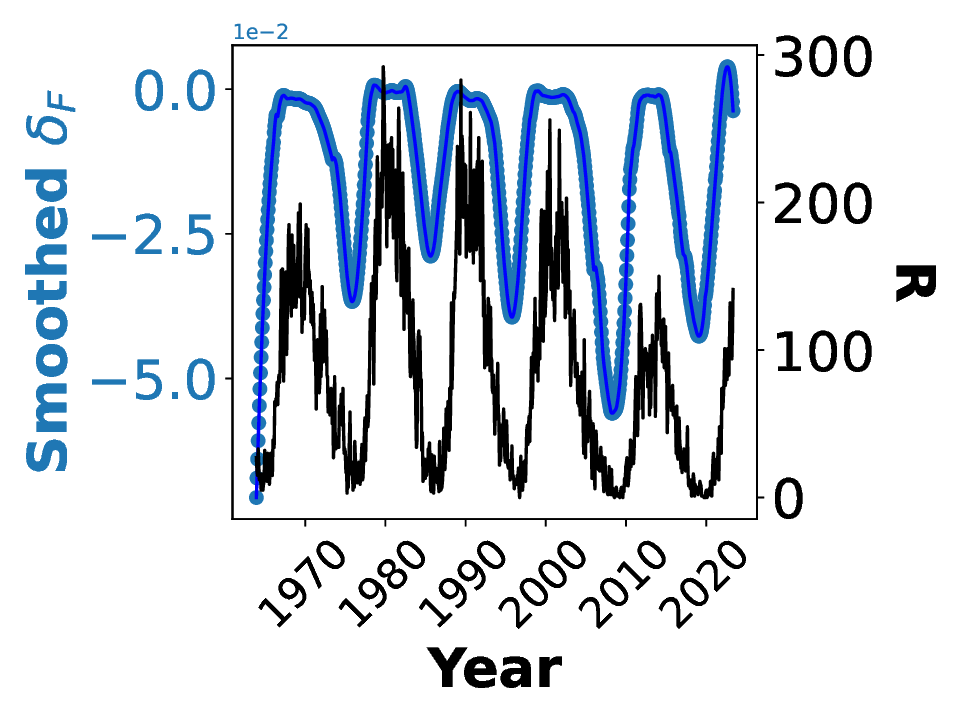}}
	\subfigure[Sunspots ]{\includegraphics[width=0.45\linewidth]{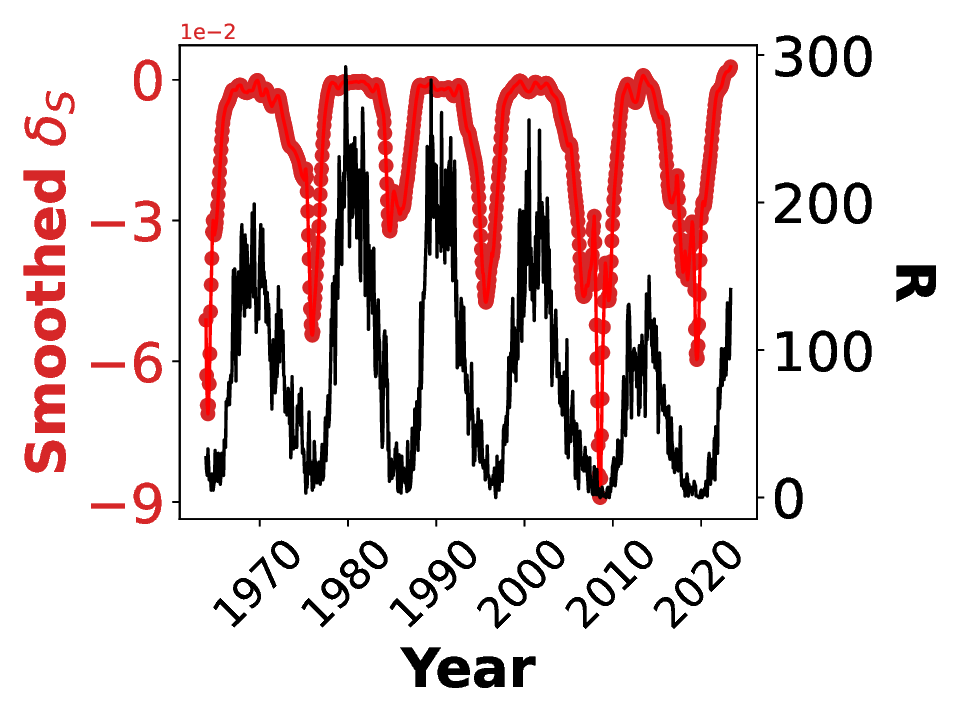}}
	\caption{Same as Fig.~\ref{fig:HVG}, but for smoothed curves
          for the $\delta$ exponent.}
	\label{fig:suavizado}
\end{figure}

Following previous literature, in order to extract possible precursor
information, we will smooth the selected parameters, the 10.7~radio flux
and the sunspots number, from Fig.~\ref{fig:HVG}, by means of a Savitzky-Golay
filter.  
The Savitzky-Golay filter, commonly used in signal processing, aims to reduce noise and improve the smoothness of signal trends. It achieves this by fitting a polynomial to each window of data, determined by the chosen polynomial degree and window size~\cite{press1990savitzky}.
 Smoothed curves are shown in 
Fig.~\ref{fig:suavizado}.

\begin{figure}[H]
	\centering
	\subfigure[F 10.7~cm index]{\includegraphics[width=0.4\linewidth]{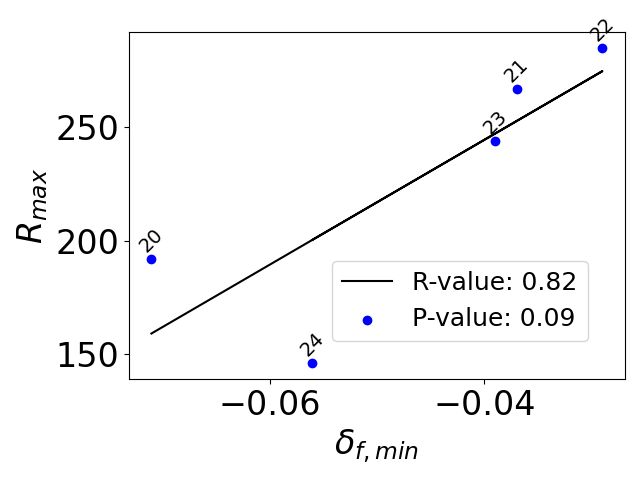}}
	\subfigure[Sunpots]{\includegraphics[width=0.4\linewidth]{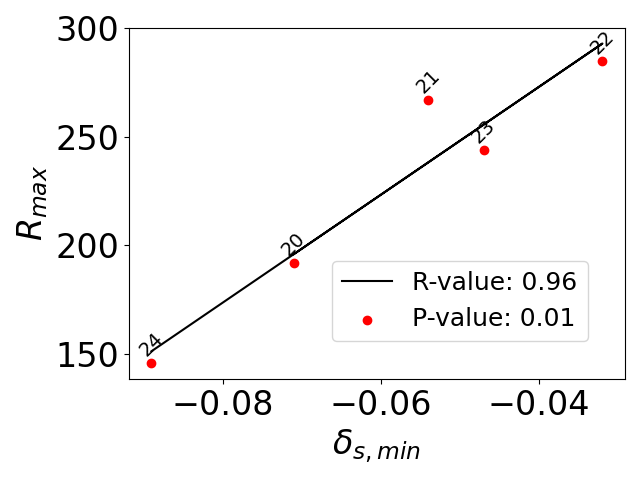}} 
	\caption{Maximum sunspots number versus minimum $\delta$ for
		each solar cycle, for solar index (blue dots)  and for sunspot (red dots).}
	\label{fig:pred_suave}
\end{figure}

Now we plot the
maximum number of sunspots in the next solar maximum ($R_\text{max}$),
versus the
 minimum value of
$\delta$ during a solar minimum ($\delta_\text{min}$), 
taken from the smoothed curves in Fig.~\ref{fig:suavizado}. Results
are shown in Fig.~\ref{fig:pred_suave}, including the corresponding
linear fits.

\begin{comment}
\begin{figure}[H]
	\centering
	\subfigure[F 10.7~cm index]{\includegraphics[width=0.4\linewidth]{pred_solarindex}}
	\subfigure[Sunpots]{\includegraphics[width=0.4\linewidth]{pred_sunspot.eps}} 
	\caption{Maximum sunspots number versus minimum $\delta$ for
          each solar cycle, for solar index (blue dots) with cc=0.57, and for sunspot (red dots) with cc=0.67.}
	\label{fig:pred_suave}
\end{figure}
\end{comment}

The linear fits are given by
\begin{align}
R_\text{max}^{\delta_{f,\text{min}}} &= 2755 \cdot \delta_{f,\text{min}} + 355 \ , \label{fits_f}\\
R_\text{max}^{\delta_{s,\text{min}}} &= 2483 \cdot \delta_{s,\text{min}} + 372 \ , \label{fits_s}
\end{align}
for the solar index and the sunspots curves, respectively.
The R-values are 0.82 for the F 10.7~cm index, and 0.96
for sunspots, whereas $p$-values is 0.09 for the solar index and 0.01 for sunspots.
Based on these results, the prediction for the maximum number of
sunspots for the next solar cycle is 
$R_\text{max}^{\delta_{f,\text{min}}}= 160$ if the solar index curve is considered, and
$R_\text{max}^{\delta_{s,\text{min}}}= 179$ if the sunspots curve is considered.

We have used this method for the previous cycles in order to obtain an estimation of the accuracy of the proposed method. 
If the fits \eqref{fits_f} and \eqref{fits_s} hold, then
Table~\ref{tab_comparacion} shows the predicted maximum number of
sunspots, along with the official number.

\begin{table}[H]
	\centering
	\caption{For each solar cycle (SC) considered, the maximum number of
          sunspots $R_{\text{max}}$ according to the Solar Cycle
          Prediction Panel (SCPP), and the results given by our
          Eqs.~\eqref{fits_f} and~\eqref{fits_s}. }
	\begin{tabular}{|c|c|c|c|c|c|c|c|}
		\hline
		\textbf{SC} & \textbf{$R_\text{max}$} & \textbf{$  \delta_{f,\text{min}} $} & \textbf{$ R_\text{max}^{\delta_{f,\text{min}}} $} & \textbf{Error \%} & \textbf{$ \delta_{s,\text{min}}$} & \textbf{$ R_\text{max}^{\delta_{s,\text{min}}} $} & \textbf{Error \%} \\
		\hline
		20 & 157 & $-0.071$ & 159 & 0.6 & $-0.071$ & 196 & 1.3 \\
		\hline
		21 & 233 & $-0.037$ & 253 & 23.9 & $-0.054$ & 238 & 7.9 \\
		\hline
		22 & 213 & $-0.029$ & 275 & 5.8 & $-0.032$ & 293 & 22.5 \\
		\hline
		23 & 180 & $-0.039$ & 248 & 10 & $-0.047$ & 255 & 27.4 \\
		\hline
		24 & 116 & $-0.056$ & 201 & 8.7 & $-0.089$ & 151 & 42.3 \\
		\hline
		25 & - & $-0.043$ & 160 & - & $-0.059$ & 179 & - \\
		\hline
	\end{tabular}
	\label{tab_comparacion}
\end{table}

Since this strategy consists
of correlating the values of one parameter during solar minimum, to
the values of another parameter during the following solar maximum,
each cycle provides one point, which affects the statistics. A similar
issue is found in 
related works. For instance, four points are also used in
Refs.~\citep{benson2020forecasting, petrovay2020solar, brown1976determines}, where the
Polar Field Strength is used.
Consequently, correlation coefficients tend to be low. In
Ref.~\citep{petrovay2020solar}, a $\mathrm{cc}=~0.676$ for the correlation between
sunspots number during solar minimum and during solar maximum, and
Ref.~ \citep{nandy2021progress} 
has $\mathrm{cc}=0.72$ for the correlation between magnetic flux at
the poles during solar minima, and the amplitude of the next solar
maximum. 

Statistics can be improved by considering previous solar cycles, but this
is possible only for the sunspots number. As mentioned above, we are
interested in comparing with recent work using the same parameters as
us, which sets the time window to be studied.

The Solar Cycle Prediction Panel predicts that the solar maximum would
be in July 2025 ($\pm$ 8 months), reaching a maximum number of
sunspots of 115. They predicted that this would be a below-average solar cycle. 
Our result of $R_\text{max}^{\delta_{s,\text{min}}}$ = 179 is closer to the behavior of the current solar cycle than the 115 predicted by the Panel.
Our result is consistent with what is mentioned in
Ref.~\citep{mcintosh2020overlapping}, where they identified the
so-called ``termination'' events that mark the end of each cycle.
Based on this, they extracted a relationship between the temporal
spacing of the termination events and the magnitude of the
cycles~\citep{leamon2020timing}, leading to the conclusion that Solar
Cycle 25 should have a magnitude greater than that proposed by the
Solar Cycle Prediction Panel. 
On the other hand, in Ref.~\citep{li2015predicting}, $R_{\text{max}}=
175$ is proposed, which is also close to what we obtain.

We can also estimate the date for the next solar maximum,
 based on previous work by Hathaway~\citep{hathaway1994shape},  where
 the number of sunspots as a function of time $R(t)$ is adjusted with 
\begin{equation}
\label{R_t}
	R(t) = \frac{ A(t-t_0)^3 }{\exp\left[ (t-t_0)^2 /B^2\right]-C}
        \ ,
\end{equation}
where $A$, $B$, $C$, are adjustable parameters, and $t_0$ is an
estimation of the time where the last solar minimum occurred. $A$,
$B$, $C$ are adjusted to fit the curve so that it starts at $R(t_0)=0$,
%\comentario{?`es cierto?. Lalo: si lo es}, 
and its maximum value is $R_\text{max}$. 
In Ref.~\cite{hathaway1994shape},
it is assumed that $t_0$ is 5.15 months
before December 2019, and adjustable parameters are $A = 0.00170 \pm 0.00040
$, $B = 14.2 + 8.33 \times A^{-1/4} $, and $C= 0.42$.

\begin{figure}[H]
	\centering
	\includegraphics[height = 0.24\textheight]{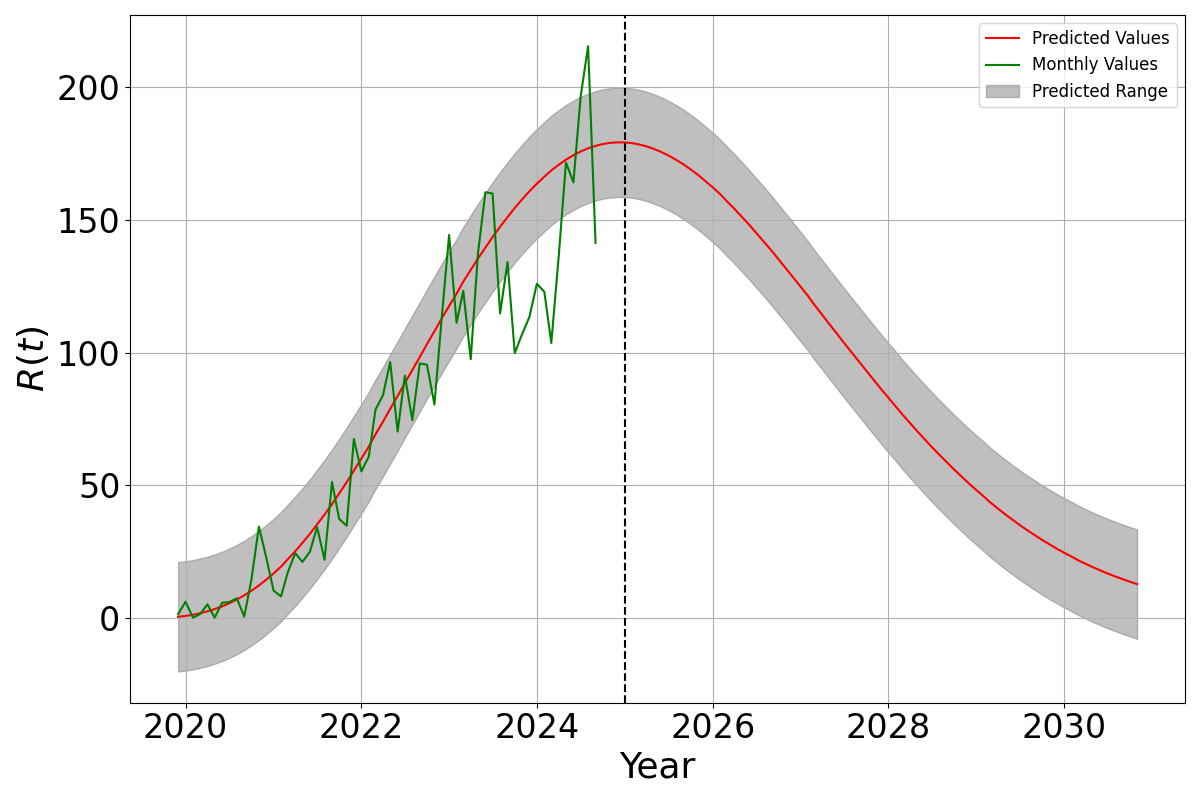}
	\caption{Sunspots number forecast for solar cycle 25. Maximum
          number of sunpots is 179, and occurs in December 2024/January 2025.}
	\label{fig:forecast}
\end{figure}

We will use the same model, Eq.~\eqref{R_t}, but will adjust
$C$ and $t_0$ to match our $R_\text{max}=179$ prediction.
Proceeding in this way, we obtain that the next solar maximum, for cyle 25, will occur in
December 2024/January 2025. Figure~\ref{fig:forecast} shows the predicted curve
Eq.~\eqref{R_t} resulting from our parameters (red), along with the
observed sunspots number for the current cycle (green), to show the
agreement between the curves so far. 

It is interesting to notice that our
results are consistent with Ref.~\cite{mcintosh2020overlapping}, but
using a different approach.

\section{Summary}
\label{summary}

We analyze the time series of five solar parameters through complex
network analysis. To do this, we construct a complex network using the
visibility graph and the horizontal visibility graph algorithms, which
convert the time series into a graph, where each node corresponds to a
data point in the series, and two nodes are connected if visibility
exists between the corresponding data points. For networks built with the
VG algorithm, the degree distribution $P(k)$  has a power-law
decay, whereas for networks built with HVG it has an exponential
decay, a result which is consistent with previous works.

A moving windows analysis show that the decay exponent $\delta$ for
the HVG exhibits a 
correlation with the solar cycle, but only for the sunspots and the F
10.7 time series: the value of $\delta$ during a minimum of the solar
cycle is lower when the number of sunspots during the following solar
maximum is higher. Fitting this to a linear trend, and using the value of
$\delta$ for the 
last solar minimum, a maximum number of sunspots of 179 is found for
the next solar maximum. Adjusting the Hathaway
function~\citep{hathaway1994shape, hathaway1999synthesis, hathaway2015solar} to reach this maximum number, 
yields that the maximum of solar cycle 25 will occur in December
2024/January 2025. 

Several improvements can be made to this approach. The same parameters
as in Ref.~\citep{singh2017early} have been used, which limits
the analysis to the last 5 solar cycles, due to data availability.
This, in turns, leads to lower statistics. As long as measurements
continue, further data can be gathered, and the quality of the
correlations could be better assessed. For sunspots, however, there
are longer records, and the method can be applied for a larger number
of solar cycles, although no comparison with other time series is
possible. Besides, in this work we have only taken the simplest
metric, the degree distribution, but various other metrics could be
considered, as previous works have shown that the clustering
coefficient, betweenness centrality, or Gini
coefficients~\citep{munoz2021analysis, munoz2022complex, zurita2023characterizing} have shown to be
useful in describing variations in solar activity. 

Previous works have proposed similar strategies, based on correlations
between physical 
parameters at solar minimum and the next solar maximum, to forecast
features of the solar cycle. Our work follows these ideas, but with a
different, complex network approach, yielding results which are
consistent with existing literature, and suggesting a new path to
explore in future research.

\section{Acknowledgments}
 We thank the Space Physics Data Facility, NASA/Goddard Space Flight Center. 

\section{Funding}
This research was funded by FONDECyT grant number 1242013 (V.M.), and supported by ANID PhD grant number
21210996 (E.F.)

%\bibliography{forecast_cycle}

\begin{thebibliography}{0}%
\makeatletter
\providecommand \@ifxundefined [1]{%
 \@ifx{#1\undefined}
}%
\providecommand \@ifnum [1]{%
 \ifnum #1\expandafter \@firstoftwo
 \else \expandafter \@secondoftwo
 \fi
}%
\providecommand \@ifx [1]{%
 \ifx #1\expandafter \@firstoftwo
 \else \expandafter \@secondoftwo
 \fi
}%
\providecommand \natexlab [1]{#1}%
\providecommand \enquote  [1]{``#1''}%
\providecommand \bibnamefont  [1]{#1}%
\providecommand \bibfnamefont [1]{#1}%
\providecommand \citenamefont [1]{#1}%
\providecommand \href@noop [0]{\@secondoftwo}%
\providecommand \href [0]{\begingroup \@sanitize@url \@href}%
\providecommand \@href[1]{\@@startlink{#1}\@@href}%
\providecommand \@@href[1]{\endgroup#1\@@endlink}%
\providecommand \@sanitize@url [0]{\catcode `\\12\catcode `\$12\catcode
  `\&12\catcode `\#12\catcode `\^12\catcode `\_12\catcode `\%12\relax}%
\providecommand \@@startlink[1]{}%
\providecommand \@@endlink[0]{}%
\providecommand \url  [0]{\begingroup\@sanitize@url \@url }%
\providecommand \@url [1]{\endgroup\@href {#1}{\urlprefix }}%
\providecommand \urlprefix  [0]{URL }%
\providecommand \Eprint [0]{\href }%
\providecommand \doibase [0]{https://doi.org/}%
\providecommand \selectlanguage [0]{\@gobble}%
\providecommand \bibinfo  [0]{\@secondoftwo}%
\providecommand \bibfield  [0]{\@secondoftwo}%
\providecommand \translation [1]{[#1]}%
\providecommand \BibitemOpen [0]{}%
\providecommand \bibitemStop [0]{}%
\providecommand \bibitemNoStop [0]{.\EOS\space}%
\providecommand \EOS [0]{\spacefactor3000\relax}%
\providecommand \BibitemShut  [1]{\csname bibitem#1\endcsname}%
\let\auto@bib@innerbib\@empty
%</preamble>
\end{thebibliography}%


\begin{thebibliography}{10}

\bibitem{schrijver2009heliophysics}
C.~J. Schrijver and G.~L. Siscoe, editors, {\em Heliophysics: {P}lasma Physics
  of the Local Cosmos\/} (Cambridge Core, 2013).

\bibitem{sturrock2013physics}
P.~A. Sturrock, {\em Physics of the {S}un. {V}olume II: {T}he Solar
  Atmosphere\/} (Springer, 2013).

\bibitem{schrijver2010heliophysics}
C.~J. Schrijver and G.~L. Siscoe, editors, {\em Heliophysics: Evolving Solar
  Activity and the Climates of Space and {E}arth\/} (Cambridge Core, 2010).

\bibitem{charbonneau2014solar}
P.~Charbonneau, Ann.\ Rev.\ Astron.\ Astrophys. {\bf 52}, 251 (2014).

\bibitem{aschwanden2006physics}
M.~Aschwanden, {\em Physics of the Solar Corona: {A}n Introduction with
  Problems and Solutions\/} (Springer, 2005).

\bibitem{kamide2007handbook}
Y.~Kamide and A.~C.-L. Chian, editors, {\em Handbook of the Solar-Terrestrial
  Environment\/} (Springer-Verlag, 2007).

\bibitem{lang2008sun}
K.~R. Lang, {\em The Sun from Space\/} (Springer, 2009).

\bibitem{sello2001solar}
S.~Sello, Astron.\ Astrophys. {\bf 377}, 312 (2001).

\bibitem{wang2017surface}
A.~R. Yeates, M.~C. Cheung, J.~Jiang, K.~Petrovay, and Y.-M. Wang, Space Sci.\
  Rev. {\bf 219}, 31 (2023).

\bibitem{jiang2014magnetic}
J.~Jiang, D.~Hathaway, R.~Cameron, S.~Solanki, L.~Gizon, and L.~Upton, Space
  Sci.\ Rev. {\bf 186}, 491 (2014).

\bibitem{cameron2017global}
R.~Cameron, M.~Dikpati, and A.~Brandenburg, Space Sci.\ Rev. {\bf 210}, 367
  (2017).

\bibitem{charbonneau2010dynamo}
P.~Charbonneau, Liv. Rev.\ Solar Phys. {\bf 2}, 2 (2005).

\bibitem{hathaway1999synthesis}
D.~H. Hathaway, R.~M. Wilson, and E.~J. Reichmann, J. Geophys.\ Res. {\bf 104},
  22375 (1999).

\bibitem{petrovay2020solar}
K.~Petrovay, Liv. Rev.\ Solar Phys. {\bf 17}, 2 (2020).

\bibitem{mcnish1949prediction}
A.~G. McNish and J.~V. Lincoln, EOS, Trans.\ Am.\ Geophys.\ Union {\bf 30}, 673
  (1949).

\bibitem{hathaway1994shape}
D.~H. Hathaway, R.~M. Wilson, and E.~J. Reichmann, Solar Phys. {\bf 151}, 177
  (1994).

\bibitem{wilson1998estimate}
R.~M. Wilson, D.~H. Hathaway, and E.~J. Reichmann, J. Geophys.\ Res. {\bf 103},
  6595 (1998).

\bibitem{feynman1982geomagnetic}
J.~Feynman, J. Geophys.\ Res. {\bf 87}, 6153 (1982).

\bibitem{lacasa2008time}
L.~Lacasa, B.~Luque, F.~Ballesteros, J.~Luque, and J.~C. Nuno, Proc.\ Nat.\
  Acad.\ Sci. {\bf 105}, 4972 (2008).

\bibitem{munoz2021analysis}
V.~Mu{\~n}oz and N.~E. Garc\'es, PLoS ONE {\bf 16}, e0259735 (2021).

\bibitem{zou2014long}
Y.~Zou, R.~V. Donner, N.~Marwan, M.~Small, and J.~Kurths, Nonlinear Proc.\
  Geophys. {\bf 21}, 1113 (2014).

\bibitem{zurita2023characterizing}
T.~Zurita-Valencia and V.~Mu{\~n}oz, Entropy {\bf 25}, 342 (2023).

\bibitem{suyal2014visibility}
V.~Suyal, A.~Prasad, and H.~P. Singh, Solar Phys. {\bf 289}, 379 (2013).

\bibitem{yu2012multifractal}
Z.~G. Yu, V.~Anh, R.~Eastes, , and D.-L. Wan, Nonlinear Proc.\ Geophys. {\bf
  19}, 657 (2012).

\bibitem{taran2022complex}
S.~Taran, E.~Khodakarami, and H.~Safari, Adv.\ Space Res. {\bf 70}, 2541
  (2022).

\bibitem{hathaway2015solar}
D.~H. Hathaway, Liv. Rev.\ Solar Phys. {\bf 7}, 1 (2010).

\bibitem{singh2017early}
A.~K. Singh and A.~Bhargawa, ass {\bf 362}, 199 (2017).

\bibitem{maguire2020evolution}
C.~A. Maguire, E.~P. Carley, J.~McCauley, and P.~T. Gallagher, Astron.\
  Astrophys. {\bf 633}, A56 (2020).

\bibitem{hartigan2015new}
P.~Hartigan and A.~Wright, Astrophys.\ J. {\bf 811}, 12 (2015).

\bibitem{nakano1968evidence}
G.~H. Nakano and H.~H. Heckman, Phys.\ Rev.\ Lett. {\bf 20}, 806 (1968).

\bibitem{williams1969proton}
D.~J. Williams and C.~O. Bostrom, J. Geophys.\ Res. {\bf 74}, 3019 (1969).

\bibitem{tapping201310}
K.~F. Tapping, Space Weather {\bf 11}, 394 (2013).

\bibitem{ohl1966wolf}
A.~I. Ohl, Soln.\ Dannye {\bf 12}, 84 (1966).

\bibitem{schatten1996solar}
K.~Schatten, D.~J. Myers, and S.~Sofia, Geophys.\ Res.\ Lett. {\bf 23}, 605
  (1996).

\bibitem{aschwanden201625}
M.~J. Aschwanden, {\em et~al.\/}, Space Sci.\ Rev. {\bf 198}, 47 (2016).

\bibitem{Dendy}
R.~O. Dendy, S.~C. Chapman, and M.~Paczuski, Plasma Phys.\ Controlled Fusion
  {\bf 49}, A95 (2007).

\bibitem{dominguez2014temporal}
M.~Dom{\'{\i}}nguez, V.~Mu{\~n}oz, and J.~A. Valdivia, J. Geophys.\ Res. {\bf
  119}, 3585 (2014).

\bibitem{Munoz_ao}
V.~Mu{\~n}oz, M.~Dom{\'{\i}}nguez, J.~A. Valdivia, S.~Good, G.~Nigro, and
  V.~Carbone, Nonlinear Proc.\ Geophys. {\bf 25}, 207 (2018).

\bibitem{zou2014complex}
Y.~Zou, M.~Small, Z.~Liu, and J.~Kurths, New J. Phys. {\bf 16}, 013051 (2014).

\bibitem{lu2018complex}
S.~Lu, H.~Zhang, X.~Li, Y.~Li, C.~Niu, X.~Yang, , and D.~Liu, Nonlinear Proc.\
  Geophys. {\bf 25}, 233 (2018).

\bibitem{abe2004scale}
S.~Abe and N.~Suzuki, Europhys.\ Lett. {\bf 65}, 581 (2004).

\bibitem{abe2006complex}
S.~Abe and N.~Suzuki, Nonlinear Proc.\ Geophys. {\bf 13}, 145 (2006).

\bibitem{abe2011finite}
S.~Abe, D.~Past{\'e}n, and N.~Suzuki, Physica A {\bf 390}, 1343 (2011).

\bibitem{pasten2017time}
D.~Past{\'e}n, F.~Torres, B.~Toledo, V.~Mu{\~n}oz, J.~Rogan, and J.~A.
  Valdivia, Pure Appl.\ Geophys. {\bf 173}, 2267 (2016).

\bibitem{abe2011universalities}
S.~Abe, D.~Past{\'e}n, V.~Mu{\~n}oz, and N.~Suzuki, Chinese Science Bulletin
  {\bf 56}, 3697 (2011).

\bibitem{pasten2018time}
D.~Past{\'e}n, Z.~Czechowski, and B.~Toledo, Chaos {\bf 28}, 083128 (2018).

\bibitem{pasten2018non}
D.~Past{\'e}n, F.~Torres, B.~Toledo, V.~Mu{\~n}oz, and J.~A. Valdivia, Physica
  A {\bf 491}, 445 (2018).

\bibitem{charakopoulos2018dynamics}
A.~K. Charakopoulos, G.~A. Katsouli, and T.~E. Karakasidis, Physica A {\bf
  495}, 436 (2018).

\bibitem{albert2002statistical}
R.~Albert and A.-L. Barab{\'a}si, Rev.\ Mod.\ Phys. {\bf 74}, 47 (2002).

\bibitem{acosta2021applying}
B.~Acosta-Tripailao, D.~Past{\'e}n, and P.~S. Moya, Entropy {\bf 23}, 470
  (2021).

\bibitem{daei2017complex}
F.~Daei, H.~Safari, and N.~Dadashi, Astrophys.\ J. {\bf 845}, 36 (2017).

\bibitem{munoz2022complex}
V.~Mu{\~n}oz and E.~Fl{\'a}ndez, Entropy {\bf 24}, 753 (2022).

\bibitem{najafi2020solar}
A.~Najafi, A.~H. Darooneh, A.~Gheibi, and N.~Farhang, Astrophys.\ J. {\bf 894},
  66 (2020).

\bibitem{gheibi2017solar}
A.~Gheibi, H.~Safari, and M.~Javaherian, Astrophys.\ J. {\bf 847}, 115 (2017).

\bibitem{lotfi2020ultraviolet}
N.~Lotfi, M.~Javaherian, B.~Kaki, A.~H. Darooneh, and H.~Safari, Chaos {\bf
  30}, 043124 (2020).

\bibitem{newman2018networks}
M.~E.~J. Newman, Phys.\ Rev.\ E {\bf 95}, 108701 (2005).

\bibitem{luque2009horizontal}
B.~Luque, L.~Lacasa, F.~Ballesteros, and J.~Luque, Phys.\ Rev.\ E {\bf 80},
  046103 (2009).

\bibitem{lacasa2009visibility}
L.~Lacasa, B.~Luque, J.~Luque, and J.~C. Nu\mbox{\~n}o, Europhys.\ Lett. {\bf
  86}, 30001 (2009).

\bibitem{kundu2021extracting}
S.~Kundu, A.~Opris, Y.~Yukutake, and T.~Hatano, Frontiers Phys. {\bf 9}, 656310
  (2021).

\bibitem{li2023convolutional}
R.~Li, W.~Li, K.~Yue, and Y.~Li, Biomed.\ Signal Proc.\ Control {\bf 86},
  104966 (2023).

\bibitem{lacasa2012time}
L.~Lacasa, A.~Nu{\~n}ez, {\'E}.~Rold{\'a}n, J.~M.~R. Parrondo, and B.~Luque,
  Eur.\ Phys.\ J. B {\bf 85}, 217 (2012).

\bibitem{telesca2020analysis}
L.~Telesca, D.~Past{\'e}n, and V.~Mu{\~n}oz, Pure Appl.\ Geophys. {\bf 177},
  4755 (2020).

\bibitem{barabasi1999emergence}
A.-L. Barab{\'a}si and R.~Albert, Science {\bf 286}, 509 (1999).

\bibitem{Guillier_b}
S.~Guillier, V.~Mu{\~n}oz, J.~Rogan, R.~Zarama, and J.~A. Valdivia, Physica A
  {\bf 467}, 465 (2016).

\bibitem{kaki2022evidence}
B.~Kaki, N.~Farhang, and H.~Safari, SIAM Review {\bf 12}, 16835 (2022).

\bibitem{press1990savitzky}
W.~H. Press and S.~A. Teukolsky, Comput.\ Phys. {\bf 4}, 669 (1990).

\bibitem{benson2020forecasting}
B.~Benson, W.~D. Pan, A.~Prasad, G.~A. Gary, and Q.~Hu, Solar Phys. {\bf 295},
  65 (2020).

\bibitem{brown1976determines}
G.~M. Brown, Mon.\ Not.\ R. Astron.\ Soc. {\bf 174}, 185 (1976).

\bibitem{nandy2021progress}
D.~Nandy, Solar Phys. {\bf 296}, 54 (2021).

\bibitem{mcintosh2020overlapping}
S.~W. McIntosh, S.~Chapman, R.~J. Leamon, R.~Egeland, and N.~W. Watkins, Solar
  Phys. {\bf 295}, 163 (2020).

\bibitem{leamon2020timing}
R.~J. Leamon, S.~W. McIntosh, S.~C. Chapman, and N.~W. Watkins, Solar Phys.
  {\bf 295}, 36 (2020).

\bibitem{li2015predicting}
K.~Li, W.~Feng, and F.~Li, J. Atmos.\ Sol.\ Terr.\ Phys. {\bf 135}, 72 (2015).

\end{thebibliography}
%\bibliographystyle{physrev}

\end{document}